\documentclass[aps,
12pt,
final,
notitlepage,
oneside,
onecolumn,
nobibnotes,nofootinbib,superscriptaddress,noshowpacs]
{revtex4-2}

\usepackage{amsmath,amssymb}
\usepackage[utf8]{inputenc}
\usepackage[english]{babel}
\usepackage{hyperref}
\usepackage{graphicx}

\begin{document}


\title{A Multi-Level Parallel Pipeline for SPHERE-3 Detector Simulation: From EAS Generation to Image Approximation\protect\\
}

\author{\firstname{V.~A.}~\surname{Ivanov}}
\email[E-mail: ]{ivanov.va18@physics.msu.ru}
\affiliation{Skobeltsyn Istitute for Nuclear Physics, Lomonosov Moscow State University, Lenenkie gory, 1(2), Moscow, 119234, Russia}
\affiliation{Faculty of Physics, Lomonosov Moscow State University, Lenenkie gory, 1(2), Moscow, 119234, Russia}

\author{\firstname{V.I.}~\surname{Galkin}}
\email[E-mail: ]{v_i_galkin@mail.ru}
\affiliation{Skobeltsyn Istitute for Nuclear Physics, Lomonosov Moscow State University, Lenenkie gory, 1(2), Moscow, 119234, Russia}
\affiliation{Faculty of Physics, Lomonosov Moscow State University, Lenenkie gory, 1(2), Moscow, 119234, Russia}

\author{\firstname{E.A.}~\surname{Bonvech}}
\email[E-mail: ]{bonvech@yandex.ru}
\affiliation{Skobeltsyn Istitute for Nuclear Physics, Lomonosov Moscow State University, Lenenkie gory, 1(2), Moscow, 119234, Russia}

\author{\firstname{O.V.}~\surname{Cherkesova}}
\affiliation{Skobeltsyn Istitute for Nuclear Physics, Lomonosov Moscow State University, Lenenkie gory, 1(2), Moscow, 119234, Russia}
\affiliation{Department of Cosmic Research, Lomonosov Moscow State University, Lenenkie gory, 1(52), Moscow, 119234, Russia}

\author{\firstname{D.V.}~\surname{Chernov}}
\affiliation{Skobeltsyn Istitute for Nuclear Physics, Lomonosov Moscow State University, Lenenkie gory, 1(2), Moscow, 119234, Russia}

\author{\firstname{T.A.}~\surname{Kolodkin}}
\affiliation{Skobeltsyn Istitute for Nuclear Physics, Lomonosov Moscow State University, Lenenkie gory, 1(2), Moscow, 119234, Russia}
\affiliation{Faculty of Physics, Lomonosov Moscow State University, Lenenkie gory, 1(2), Moscow, 119234, Russia}

\author{\firstname{N.O.}~\surname{Ovcharenko}}
\affiliation{Skobeltsyn Istitute for Nuclear Physics, Lomonosov Moscow State University, Lenenkie gory, 1(2), Moscow, 119234, Russia}
\affiliation{Faculty of Physics, Lomonosov Moscow State University, Lenenkie gory, 1(2), Moscow, 119234, Russia}

\author{\firstname{D.A.}~\surname{Podgrudkov}}
\affiliation{Skobeltsyn Istitute for Nuclear Physics, Lomonosov Moscow State University, Lenenkie gory, 1(2), Moscow, 119234, Russia}
\affiliation{Faculty of Physics, Lomonosov Moscow State University, Lenenkie gory, 1(2), Moscow, 119234, Russia}

\author{\firstname{T.M.}~\surname{Roganova}}
\affiliation{Skobeltsyn Istitute for Nuclear Physics, Lomonosov Moscow State University, Lenenkie gory, 1(2), Moscow, 119234, Russia}

\author{\firstname{M.~D.}~\surname{Ziva}}
\affiliation{Skobeltsyn Istitute for Nuclear Physics, Lomonosov Moscow State University, Lenenkie gory, 1(2), Moscow, 119234, Russia}
\affiliation{Faculty of Computational Mathematics and Cybernetics, Lomonosov Moscow State University, Lenenkie gory, 1(52), Moscow, 119234, Russia}



\received{June 13, 2018} 

\begin{abstract}
Optimization of the SPHERE-3 detector configuration, designed to study the mass composition of primary cosmic rays in the energy range 1--1000~PeV by registering Cherenkov light reflected from the snow surface, requires simulation of a large number of extensive air shower events. A software suite with a multi-step computational pipeline is presented: shower generation in CORSIKA, decoding and cloning of events (C++/OpenMP), ray-tracing of optical photons through the detector model (Geant4~MT), and approximation of images by a lateral distribution function (Python/multiprocessing, iminuit). The key property of the problem is its natural atomicity: each event is processed independently at all stages, which provides linear scaling under parallel computation. Thread safety is achieved by architectural means --- shared data are read-only, mutable state is isolated per-worker --- without the use of locks on hot paths.
\end{abstract}



\maketitle

\section{Introduction}
\label{sec:intro}

The problem of detecting primary cosmic radiation (PCR) has been actively addressed for many decades. Recent results~\cite{thoudam} indicate that the majority of primary cosmic ray (PCR) events with energies of 1--1000~PeV may be of extragalactic origin. Thus, the composition of PCR in this energy range may be of crucial importance for constructing a model of the transition from galactic to extragalactic cosmic rays. Understanding the physics of this process is essential for describing the acceleration and propagation of cosmic rays.

The detection method based on registering Cherenkov light (CL) from extensive air showers (EAS) reflected off the snow surface was proposed by A.\,E.~Chudakov~\cite{chudakov} and became the basis for the series of SPHERE experiments. The results obtained with the SPHERE-2 detector~\cite{antonov} confirmed the promise of the method and provided the foundation for developing the next-generation detector --- SPHERE-3~\cite{chernov}. The SPHERE-3 reflected CL telescope  will feature a substantially improved light-collecting power (effective entrance window area of at least 1~m$^2$) and an optical resolution of about 2000~pixels owing to the use of a modified Schmidt optical system with a SiPM photodetector mosaic~\cite{sipm}.

Another novelty of the SPHERE-3 setup is the abandoning of the balloon carrier in favor of the UAV, thus, one can observe the upper hemisphere. Namely, a detector of the angular distribution of direct CL will be used that can help to accurately estimate the shower direction and contribute much to the primary nucleus mass definition.

Optimization of the detector configuration and development of criteria for shower classification by primary particle mass require accumulation of large statistics of simulated events. The full parameter space of the simulation includes 5~types of primary nuclei, 5~energy values, 5~zenith angles, several EAS models, and atmospheric models. Taking into account the cloning procedure --- augmentation of statistics by parallel translation of the shower axis --- and subsequent photon tracing, the total number of processed events reaches the order of $10^6$. The key property of the problem is its \textit{natural atomicity}: each event at all processing stages is independent of the others, which allows efficient parallelization of computations without loss of physical meaning.

\begin{figure}[htbp]
\centering
\includegraphics[width=0.95\textwidth]{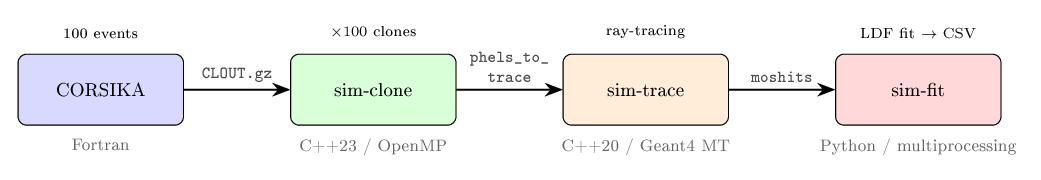}
\caption{Computational pipeline for the SPHERE-3 detector simulation. Intermediate data formats and parallelism technologies are indicated for each stage.}
\label{fig:pipeline}
\end{figure}

In this work, a software suite for the full simulation of the SPHERE-3 reflected CL telescope
is presented (Fig.~\ref{fig:pipeline}). The pipeline consists of four stages: generation of EAS events in CORSIKA~\cite{corsika} (Section~\ref{sec:corsika}), decoding of binary data and cloning of events using C++/OpenMP (Section~\ref{sec:sim-clone}), ray-tracing of optical photons through the geometric model of the detector based on Geant4~\cite{geant4} (Section~\ref{sec:g4}), and approximation of the registered images by a lateral distribution function using Python/multiprocessing (Section~\ref{sec:appro}). In Section~\ref{sec:discussion}, the parallelism architecture of the entire suite is discussed. As the SPHERE-3 setup also includes a detector aimed at the direct CL angular images, Section 7 contains a few lines about its capabilities.

\section{EAS event generation in CORSIKA}\label{sec:corsika}

The first stage of the computational pipeline (Fig.~\ref{fig:pipeline}) is the simulation of extensive air showers in CORSIKA~\cite{corsika}. For each set of physical parameters (primary particle type, energy, arrival angles, EAS model, and atmospheric model), a base sample of 100 events is generated, which is sufficient for analyzing the main fluctuations of longitudinal shower development. CORSIKA takes into account the spectral sensitivity of photosensors and light absorption in the atmosphere, which makes it possible to optimize the calculations and track only those photons that can be registered by the detectors. Instead of using the standard CORSIKA output files, which possess excessive detail and large size, the data are stored as multidimensional arrays corresponding to the required level of detail.

\subsection{Simulation parameters}\label{subsec:corsika-params}

The simulation is performed on discrete sets of primary particle parameter shown in Table~\ref{tab:corsika-params}. Azimuthal angles are distributed randomly (uniformly) in the range $0\text{--}360^{\circ}$.

\begin{table}[htbp]
\centering
\caption{CORSIKA simulation parameters.}
\label{tab:corsika-params}
\begin{tabular}{lcl}
\hline
Parameter &\phantom{000}& Values \\
\hline
Primary energy && 1, 3, 10, 30, 100~PeV \\
Zenith angle && $5^{\circ}$, $10^{\circ}$, $15^{\circ}$, $20^{\circ}$, $25^{\circ}$ \\
Nucleus type && H, He, N, S, Fe \\
EAS model && QGSJET01~\cite{qgsjet01}, QGSJETII-04~\cite{qgsjet2} \\
Atmospheric model && several from the CORSIKA set \\
Base sample && 100 events per set of parameters \\
\hline
\end{tabular}
\end{table}

The full parameter space comprises: 5 energies $\times$ 5 zenith angles $\times$ 5 nucleus types $\times$ $N_{\mathrm{atm}}$ atmospheric models $\times$ $N_{\mathrm{mod}}$ EAS models. Taking into account the subsequent cloning (Section~\ref{sec:sim-clone}), each base event produces up to 100 images on the detector entrance window, yielding samples of the order of $10^4$~events per grid point. The grid parameters can be adjusted and extended in the course of the work.

\subsection{Output data format}\label{subsec:corsika-format}

For each event, the spatio-temporal distribution of Cherenkov light at the snow surface level is stored. The data are represented as a three-dimensional array on a $3.2 \times 3.2$~km$^2$ grid divided into $1280 \times 1280$ cells of size $2.5 \times 2.5$~m$^2$. The time pulse of light in each cell is recorded as an array of 102 elements with a 5~ns step (100 data layers and 2 service layers). The data are stored in Fortran unformatted sequential access binary format (\texttt{CLOUT*.gz}), compressed with gzip. The size of a single array before compression is

\begin{equation}\label{eq:clout-size}
V = 1280 \times 1280 \times 102 \times 4~\text{bytes} \approx 670~\text{MB}.
\end{equation}

In addition to the distribution on the snow surface, for prospective analysis of direct Cherenkov light, photon distributions at three observation levels ($h = 0.5$; $1.0$, and $2.0$~km above the snow) are stored in arrays of dimension $40 \times 40 \times 50 \times 50 \times 30$, where the first two dimensions correspond to spatial coordinates ($400 \times 400$~m$^2$), the next two to angular coordinates ($50^{\circ} \times 50^{\circ}$), and the last to time delay (60~ns). Joint registration of reflected and direct light makes it possible to substantially improve the accuracy of cosmic ray mass composition estimation.

Each parameter set is simulated independently, which allows running CORSIKA in parallel without inter-process coordination.

\section{EAS Event Decoding and Cloning (sim-clone)}\label{sec:sim-clone}

The second stage of the pipeline converts CORSIKA output data into a set of photoelectron files on the detector entrance window. This task is performed by the \texttt{sim-clone} application, implemented in C++23 with OpenMP for parallel computation.

\subsection{Physical Meaning of Cloning}\label{subsec:cloning}

\begin{figure}[htbp]
\centering
\includegraphics[width=0.85\textwidth]{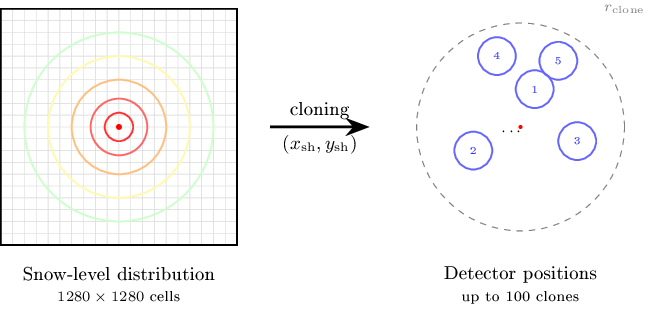}
\caption{Cloning scheme: a single Cherenkov light distribution on the snow surface (left) generates multiple images on the detector entrance window at various offsets $(x_{\mathrm{sh}}, y_{\mathrm{sh}})$ within a circle of radius $r_{\mathrm{clone}}$ (right).}
\label{fig:cloning}
\end{figure}

A single EAS event produces different Cherenkov light images on the detector entrance window depending on the relative position of the shower axis and the optical axis of the telescope. To increase the statistical sample without re-running the computationally expensive CORSIKA simulation, a cloning procedure is applied (Fig.~\ref{fig:cloning}): the original Cherenkov light distribution on the snow is translated by a random vector $(x_{\mathrm{sh}},\, y_{\mathrm{sh}})$, generated uniformly within a circle of radius $r_{\mathrm{clone}}$:
\begin{equation}\label{eq:cloning}
x_{\mathrm{sh}} = r_{\mathrm{clone}} \sqrt{\xi_1} \cos(2\pi\xi_2), \qquad y_{\mathrm{sh}} = r_{\mathrm{clone}} \sqrt{\xi_1} \sin(2\pi\xi_2),
\end{equation}
where $\xi_1, \xi_2 \sim U(0,1)$ are uniformly distributed random numbers. The cloning radius is $r_{\mathrm{clone}} = 500$~m for detector elevation heights $h \geq 600$~m and $r_{\mathrm{clone}} = 300$~m for $h < 600$~m.

Up to 100 clones are generated from each base event, yielding $10^4$ statistically distinct images on the entrance window for a typical sample of 100 CORSIKA showers. Photons falling outside the detector field of view ($\sqrt{x^2 + y^2} > r_{\max}$, where $r_{\max} = 400$~m for $h \geq 600$~m) are discarded at this stage.

\subsection{CORSIKA Data Processing}\label{subsec:sim-clone-arch}

The CORSIKA output files (\texttt{CLOUT4w\_*.gz}) are gzip-compressed binary files in Fortran unformatted sequential access format. The data contain an event header and a three-dimensional array of Cherenkov light intensity of size $N_t \times N_y \times N_x = 102 \times 1280 \times 1280$ elements of type \texttt{float32} (spatial resolution 2.5~m, temporal resolution 5~ns, volume $\approx 670$~MB per event). Reading is performed by a streaming gzip decoder (zlib) without prior decompression to disk.

After loading the array, an index of nonzero elements is built in parallel; the number of such elements is substantially smaller than the full array size of $1280^2 \times 102 \approx 1.67 \cdot 10^8$. This allows processing only significant cells during photon generation.

\subsection{Photoelectron Generation}\label{subsec:phel-gen}

For each nonzero cell $(i, j, k)$, the coordinates on the snow are computed as
\begin{equation}\label{eq:cell-coords}
x_{ij} = 2.5\left(i - \frac{N_x}{2} - 0.5\right) + x_{\mathrm{sh}}, \qquad y_{ij} = 2.5\left(j - \frac{N_y}{2} - 0.5\right) + y_{\mathrm{sh}},
\end{equation}
after which the distance from the detector axis is determined as $\rho = \sqrt{h^2 + x_{ij}^2 + y_{ij}^2}$. The observation angle $\alpha = \arccos(h/\rho)$ determines the angular efficiency coefficient, approximated by a third-degree polynomial \cite{Antonov2015}:
\begin{equation}\label{eq:eta-alpha}
\eta(\alpha) = \frac{(a_0 + a_1\alpha + a_2\alpha^2 + a_3\alpha^3) \cdot \cos^2\alpha}{2.8167},
\end{equation}
where the coefficients are $a_0 = 0.9952$, $a_1 = -3.571 \cdot 10^{-4}$, $a_2 = -1.309 \cdot 10^{-5}$, $a_3 = -3.571 \cdot 10^{-7}$ (angle $\alpha$ in degrees). The number of photoelectrons is generated according to the Poisson distribution:
\begin{equation}\label{eq:poisson-nph}
n_{\mathrm{ph}} \sim \mathrm{Pois}\left(\frac{k_{\mathrm{ph}} \cdot 1.9 \cdot I_{ijk} \cdot \eta(\alpha)}{\rho^2}\right),
\end{equation}
where $k_{\mathrm{ph}} = 1.36848$ is the quantum efficiency coefficient and $I_{ijk}$ is the intensity value in the CORSIKA array cell.

For each generated photon, the coordinates and time are subjected to uniform smearing within the cell: $\xi, \eta \sim U(-1.25, 1.25)$~m in space, $\tau \sim U(-2.5, 2.5)$~ns in time. The total photon arrival time is computed taking into account the propagation delay and shower inclination:
\begin{equation}\label{eq:photon-time}
t = t_k + \tau + t_{\mathrm{big}} + \frac{\rho}{c} + \frac{(x_i\cos\varphi + y_j\sin\varphi)\sin\theta}{c},
\end{equation}
where $t_{\mathrm{big}} = (H_{\max} - 455)/(c\cos\theta)$ is the base delay determined by the shower maximum height $H_{\max}$, and $\theta$ and $\varphi$ are the zenith and azimuthal angles of the primary particle.

\subsection{Parallelism and Data Format}\label{subsec:sim-clone-parallel}

Parallelism is implemented using OpenMP at the level of processing nonzero array cells within a single clone, with dynamic scheduling for load balancing. The atomic data unit of the pipeline is a single text file \texttt{phels\_to\_trace}, corresponding to one clone of one shower (up to 100 clones from each CLOUT file). Each file contains a header with shower parameters and records of the form $(i, j, k, m, x, y, t)$ --- cell indices, photon number, coordinates, and arrival time on the detector entrance window. Thread safety details are discussed in Section~\ref{sec:discussion}.

\section{Ray-Tracing in the Detector Model (sim-trace)}\label{sec:g4}

The third stage of the pipeline is Monte Carlo ray-tracing of optical photons through the geometric model of the SPHERE-3 detector. The application is implemented in C++20 based on the Geant4 framework~\cite{geant4} using the built-in Geant4 MT multithreading support.

\subsection{Detector Geometry Model}\label{subsec:g4-geometry}

\begin{figure}[htbp]
\centering
\includegraphics[width=0.7\textwidth]{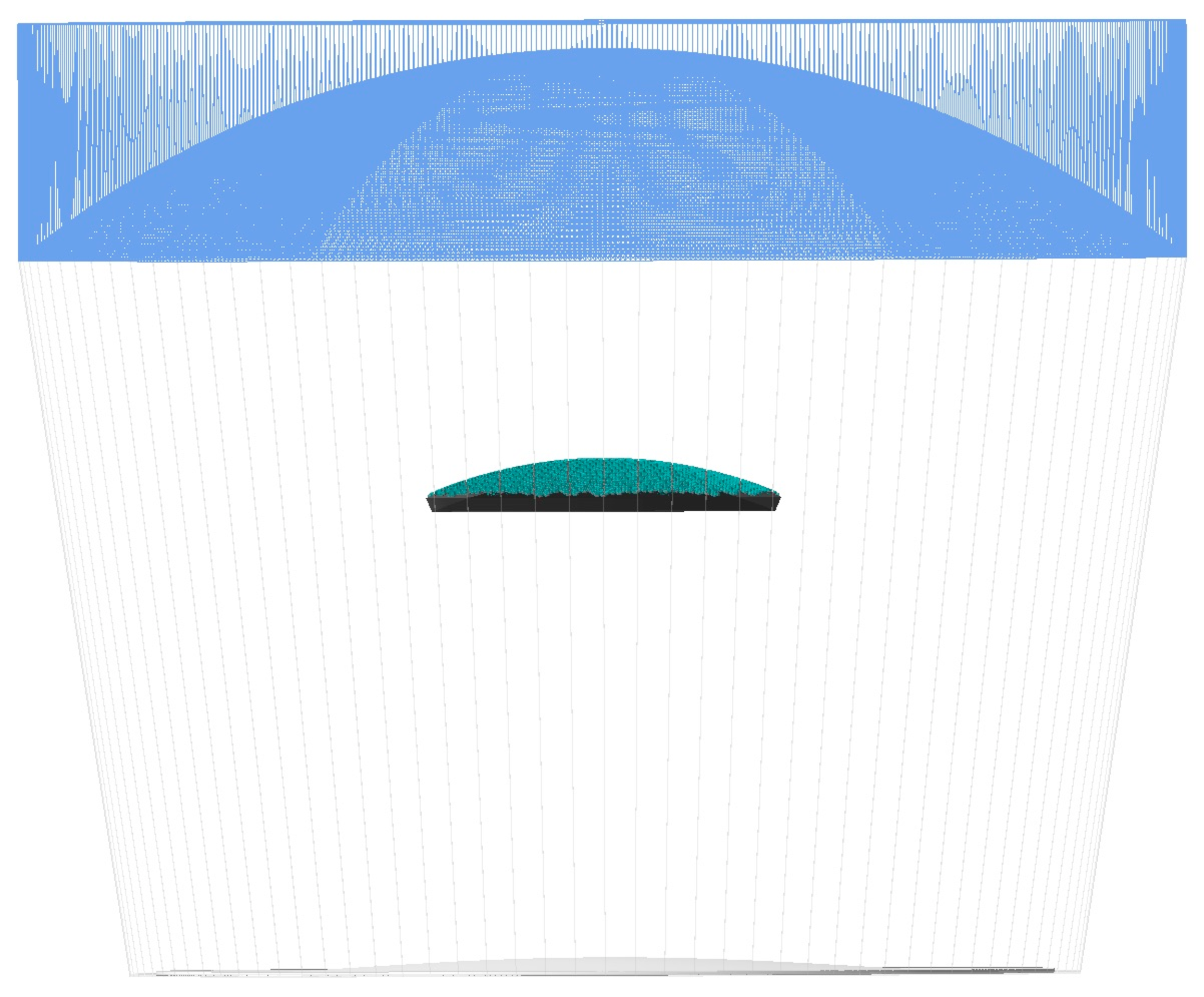}
\caption{Geant4 geometric model of the SPHERE-3 detector: spherical mirror, corrector lens, 2653-pixel SiPM mosaic, and protective hood.}
\label{fig:detector}
\end{figure}

The detector model (Fig.~\ref{fig:detector}) is placed inside a cylindrical world volume (radius 2~m, half-height 4~m, filled with air at $n = 1.00029$) and includes four main elements.

The primary optical element --- the mirror --- is imported from an STL mesh file using the CADMesh library~\cite{cadmesh}. A tessellated representation allows an accurate description of the aspherical surface without analytical parameterization. The mirror material is aluminum with specular reflection ($R = 88\text{--}85$\% in the $1.5\text{--}2.0$~eV range, modeled as a dielectric--metal boundary).

The mosaic support structure is a spherical shell ($r_{\mathrm{in}} = 83.25$~cm, $r_{\mathrm{out}} = 85.83$~cm, angular coverage $0\text{--}21^{\circ}$) made of aluminum with full absorption.

The photodetector array comprises 2653~pixels grouped in sets of~7 into 379~segments, with positions and orientations specified by a configuration file. Each pixel is modeled as a composite volume: the collector combines a hexagonal acrylic prism (radius 6~mm, height 4~mm) with a spherical drop lens (sphere radius 17~mm, $n = 1.5122$); the photocathode is a thin parallelepiped ($3 \times 3 \times 0.2$~mm) at the base of the collector.

A protective hood consisting of an aluminum ring and a conical wall connects to the mosaic; both surfaces act as full absorbers.

\subsection{Ray-Tracing Physics}\label{subsec:g4-physics}

The list of physical processes is limited to optics: boundary interaction, bulk absorption, and Rayleigh scattering. All non-optical secondary particles are killed immediately.

Each primary photon is created with energy $E = 2$~eV and random linear polarization. The entry point on the detector entrance window is generated uniformly within a circle of radius $R = 85$~cm:
\begin{equation}\label{eq:entry-point}
r = 85\text{ cm} \cdot \sqrt{\xi_1}, \qquad \varphi = 2\pi\xi_2,
\end{equation}
after which the coordinates are rotated into the detector coordinate system. The photon momentum direction is computed as the unit vector from the source point on the snow to the entry point.

When a photon crosses the photocathode volume, the cosine of the incidence angle is computed as
\begin{equation}
\cos\xi = -(\mathbf{d} \cdot \hat{\mathbf{n}}),
\end{equation}
where $\mathbf{d}$ is the momentum direction and $\hat{\mathbf{n}}$ is the pixel unit normal. If $\cos\xi > 0$, the angular sensitivity is
\begin{equation}\label{eq:sensitivity}
S(\xi) = (\cos\xi)^{p_1}, \qquad p_1 = 1.093,
\end{equation}
and the photon is registered stochastically with probability $S(\xi)$. In addition, at each detection event a spurious crosstalk hit is independently generated for each of the 6~neighboring pixels in the segment with probability $P_{\mathrm{xt}} = 0.07$.

\subsection{Multithreading Architecture}\label{subsec:g4-mt}

\begin{figure}[htbp]
\centering
\includegraphics[width=0.85\textwidth]{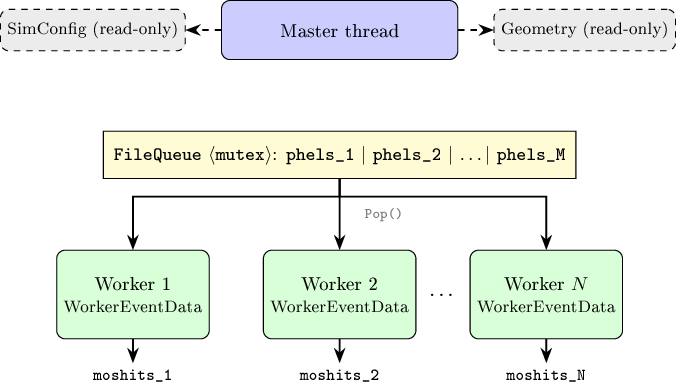}
\caption{Multithreading architecture of sim-trace. The master thread creates shared read-only geometry; worker threads dequeue files from the thread-safe \texttt{FileQueue} and write results to isolated output streams.}
\label{fig:g4mt}
\end{figure}

Multithreading is implemented using Geant4 MT (Fig.~\ref{fig:g4mt}) following the ``one event = one file'' model: each worker thread dequeues a \texttt{phels\_to\_trace} filename from the thread-safe \texttt{FileQueue} (the only synchronization point), reads all photons as primary vertices, traces them through the geometry, and writes the detection results to its own output file.

The architecture separates data into three categories: immutable configuration \texttt{SimConfig} (orientation angles, paths), shared geometry and physics tables (read-only, guaranteed by Geant4 MT), and isolated per-worker state \texttt{WorkerEventData} (photon metadata, output stream, counters). The output text files \texttt{moshits/moshits\_*} contain, for each registered photon: segment and pixel numbers, coordinates and time at the photocathode, momentum direction, origin code, and source metadata. Thread safety details are discussed in Section~\ref{sec:discussion}.

\section{Shower Image Approximation (sim-fit)}\label{sec:appro}

The final stage of the pipeline is the approximation of registered Cherenkov light images by a lateral distribution function (LDF) and the extraction of physical shower parameters. The application is implemented in Python 3.14 using \texttt{multiprocessing} for parallelization and the \texttt{iminuit} library~\cite{iminuit} for nonlinear optimization.

\subsection{Physical Model}\label{subsec:appro-model}

The radial distribution of Cherenkov light intensity on the detector mosaic is described by a four-parameter LDF model:
\begin{equation}\label{eq:ldf}
F(r) = \frac{p_0^2}{\left(1 + p_1 r\right)^2 \left(1 + p_4 r^s\right)},
\end{equation}
where $r = \sqrt{(x - x_0)^2 + (y - y_0)^2}$ is the distance from the shower axis in focal plane coordinates. The model parameters and their bounds are listed in Table~\ref{tab:ldf-params}.

\begin{table}[htbp]
\centering
\caption{LDF model parameters and their physical meaning.}
\label{tab:ldf-params}
\begin{tabular}{lll}
\hline
Parameter\phantom{000} & Description  & Bounds \\
\hline
$p_0$ & Amplitude (proportional to $\sqrt{I_{\max}}$) & $(0,\; 2\sqrt{I_{\max}})$ \\
$p_1$ & Core width (mm$^{-1}$) & $(0,\; 10^{-2})$ \\
$p_4$ & Tail amplitude & $(0,\; 10)$ \\
$s$ & Tail power index & $(0.5,\; 1.5)$ \\
$x_0, y_0$ & Shower center coordinates (mm) & $(-300,\; 300)$ \\
\hline
\end{tabular}
\end{table}

\begin{figure}[htbp]
\centering
\includegraphics[width=\textwidth]{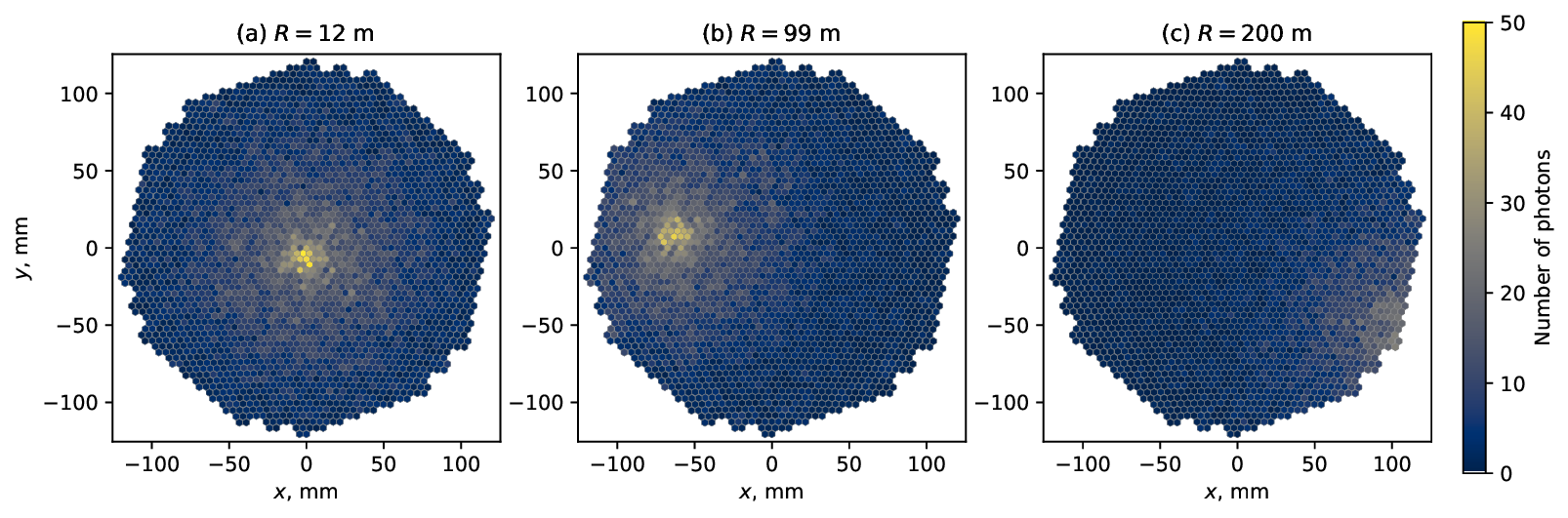}
\caption{Cherenkov light images on the SPHERE-3 detector mosaic for proton showers (10~PeV) at different distances $R$ from the detector axis. The color scale indicates the number of registered photons per pixel.}
\label{fig:mosaic}
\end{figure}

The detector pixel coordinates are converted from hardware $(x, y, z)$ to a focal plane projection:
\begin{equation}\label{eq:focal-proj}
x' = \frac{x}{z} \cdot f, \qquad y' = \frac{y}{z} \cdot f,
\end{equation}
where $f = 330$~mm is the focal length of the optical system. Before approximation, the median background level $b$, estimated from a sample of background events, is subtracted element-wise from the data: $I'_k = \max(I_k - b, 0)$. Typical Cherenkov light images on the mosaic are shown in Fig.~\ref{fig:mosaic}.

\subsection{Optimization Strategy}\label{subsec:appro-optim}

Fitting is performed by the \texttt{iminuit} library (a wrapper over CERN Minuit2~\cite{james-minuit}). Initial parameter values are determined adaptively from data statistics. The amplitude is initialized as $p_0^{(0)} = \sqrt{\max(I_{\max},\, 1)}$. The shower center coordinates are computed as the weighted mean over the $N_{\mathrm{top}} = 10$ brightest pixels:
\begin{equation}\label{eq:init-center}
x_0^{(0)} = \frac{\sum_{k=1}^{N_{\mathrm{top}}} I_k x_k}{\sum_{k=1}^{N_{\mathrm{top}}} I_k}, \qquad y_0^{(0)} = \frac{\sum_{k=1}^{N_{\mathrm{top}}} I_k y_k}{\sum_{k=1}^{N_{\mathrm{top}}} I_k},
\end{equation}
where the summation runs over pixels with the highest intensity $I_k$. The remaining parameters are initialized with fixed values: $p_1^{(0)} = 10^{-4}$~mm$^{-1}$, $p_4^{(0)} = 1.0$, $s^{(0)} = 0.7$.

\begin{figure}[htbp]
\centering
\includegraphics[width=0.7\textwidth]{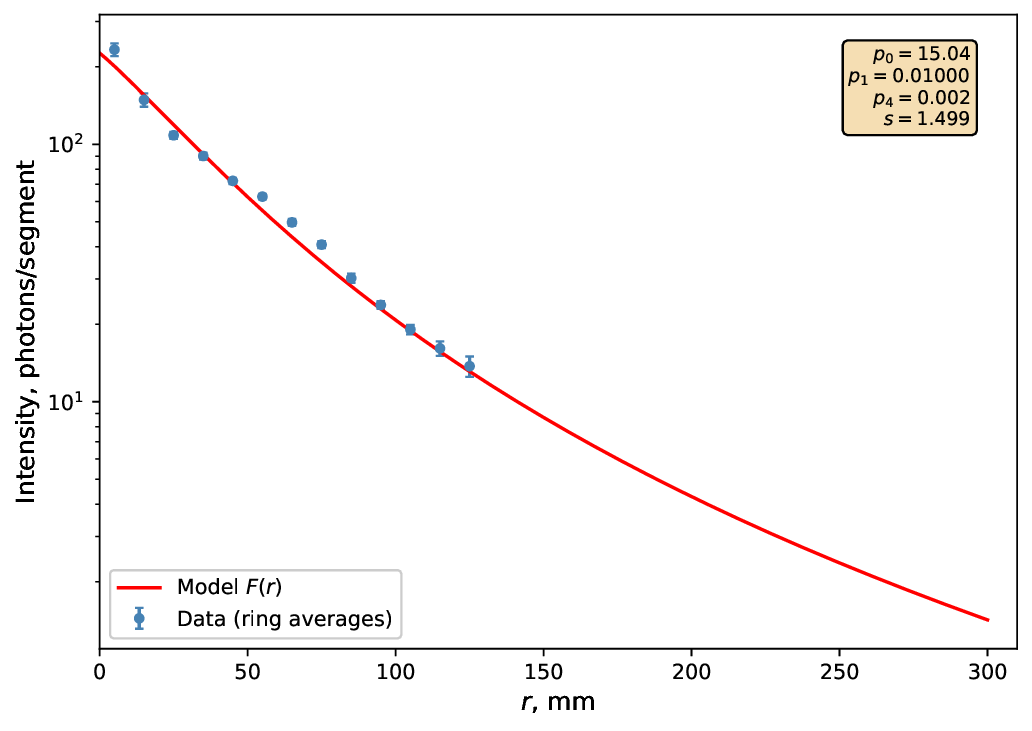}
\caption{Example LDF approximation: ring-averaged pixel intensities (points with error bars) and the fitted model $F(r)$~\eqref{eq:ldf} (solid curve).}
\label{fig:ldf}
\end{figure}

To improve robustness, the optimization is performed up to 5 times. At each restart, the initial values of $p_0$ and $p_1$ are scaled by a random factor from the range $[0.5;\; 2.0]$, and the center coordinates are shifted by a random vector $(\Delta x, \Delta y) \sim U(-20, 20)$~mm. Each attempt follows the sequence Simplex $\to$ Migrad $\to$ Hesse; the result is selected by the minimum of the regularized objective function $f_{\mathrm{reg}} = f_{\mathrm{val}} + 10^{-4}(p_1^2 + p_4^2)$. An example of an LDF approximation is shown in Fig.~\ref{fig:ldf}.

Depending on the event intensity, different statistics are used. For $I_{\max} \geq 20$, a $\chi^2$ with floor:
\begin{equation}\label{eq:chi2-floor}
\chi^2_{\mathrm{floor}} = \sqrt{\frac{1}{N}\sum_k \frac{(F_k - I_k)^2}{\max(I_k, 1)}}.
\end{equation}
For $I_{\max} < 20$, a blend statistic combining Cash statistic (Poisson likelihood) for pixels with a small number of hits and weighted $\chi^2$ for the rest:
\begin{equation}\label{eq:blend}
L_{\mathrm{blend}} = \frac{2}{N}\left[\sum_{I_k < 5}\left(F_k - I_k\ln F_k\right) + \frac{1}{2}\sum_{I_k \geq 5}\frac{(F_k - I_k)^2}{\max(I_k, 1)}\right].
\end{equation}

\subsection{LDF Integration}\label{subsec:appro-integral}

After fitting, the radial integral of the LDF is computed to estimate the total shower intensity:
\begin{equation}\label{eq:ldf-integral}
I_{\mathrm{total}} = \int_0^{330} \frac{p_0^2 \cdot r}{\left(1 + p_1 r\right)^2 \left(1 + p_4 r^s\right)} \, dr,
\end{equation}
using SciPy adaptive quadrature. The upper integration limit (330~mm) corresponds to the focal length of the optical system.

\subsection{Parallelism and Data Format}\label{subsec:appro-parallel}

Parallelism is implemented at the process level (\texttt{multiprocessing.Pool}), where the atomic unit of work is a single \texttt{moshits} file. When the pool is created, each worker loads the detector geometry (2653 pixels) once; thereafter, only the file path is passed between the master and workers. Processing is performed via \texttt{imap\_unordered} for maximum core utilization. Optionally, pixel coordinates are placed in \texttt{shared\_memory} to eliminate geometry duplication across processes.

Events are filtered by three criteria: the distance of the shower axis from the detector center ($\leq 400$~mm), the total intensity after background subtraction ($\geq I_{\min}$), and the distance of the peak from the mosaic center ($\leq 270$~mm). Results are written to CSV files (separately by particle type) containing the approximated LDF parameters, fit quality metrics, and the radial integral. Thread safety details are discussed in Section~\ref{sec:discussion}.

\section{Discussion of the Parallel Architecture}\label{sec:discussion}

The applications described in Sections~3--5 form a unified computational pipeline in which each stage employs its own parallelism model. This section discusses the common architectural principles that ensure the efficiency and correctness of parallel computations throughout the entire suite.

\subsection{End-to-End Atomicity}\label{subsec:atomicity}

The key property of the simulation problem is its natural atomicity at all levels of the pipeline. Each CORSIKA shower produces an independent Cherenkov light distribution; each clone of that shower corresponds to a statistically independent position of the shower axis; each image on the detector mosaic results from an independent ray-tracing; each LDF approximation is determined solely by the data of a single image. This chain of independence
\[
\text{shower} \rightarrow \text{clone} \rightarrow \text{mosaic image} \rightarrow \text{approximation}
\]
guarantees that no processing step for one event requires information about any other event. Consequently, within each stage there is no need for inter-process (or inter-thread) communication during data processing.

The only synchronization points in the entire pipeline are:
\begin{itemize}
\item the file system between stages, where the output of one stage serves as the input to the next;
\item a thread-safe queue inside the ray-tracing application, requiring a single mutex acquisition per event;
\item result accumulation in the fitting stage, where the master process collects outputs as workers complete.
\end{itemize}

\subsection{Hierarchy of Parallelism Models}\label{subsec:hierarchy}

Each pipeline stage employs a parallelism model tailored to the nature of its computational workload (Table~\ref{tab:parallelism}).

\begin{table}[htbp]
\centering
\caption{Parallelism models used at each pipeline stage.}
\label{tab:parallelism}
\begin{tabular}{lllll}
\hline
Stage & Technology & Level & Granularity & Shared state \\
\hline
CORSIKA & External script & Processes (OS) & One run & None \\
sim-clone & C++23/OpenMP & Threads (SHM) & Array cells & CL array (read-only) \\
sim-trace & C++20/Geant4 MT & Threads (Geant4) & One file/event & Geometry (read-only) \\
sim-fit & Python/multiprocessing & Processes & One file & Coordinates (read-only) \\
\hline
\end{tabular}
\end{table}

The choice of technology at each stage is determined by the structure of the problem. Shower simulation in CORSIKA is completely independent between runs; parallelization is trivial and is realized by an external script that distributes the jobs.

In \texttt{sim-clone}, each CLOUT file contains $\sim 10^5$--$10^6$ nonzero cells of the three-dimensional array. The computations are uniform in structure (one formula per cell with Poisson generation), which corresponds to a SIMD-like workload. OpenMP with dynamic scheduling effectively balances the nonuniformity arising from cells with higher intensity producing more photons.

In \texttt{sim-trace}, tracing a single photon through the detector geometry requires substantial computation (intersections with tessellated surfaces, multiple reflections and refractions). Geant4~MT natively supports event-level parallelism, where each event is processed by a separate thread. The geometry and physics tables are shared among threads in read-only mode, as guaranteed by the Geant4 architecture.

In \texttt{sim-fit}, nonlinear optimization involves calls to the Python interpreter and NumPy operations. The use of processes instead of threads is dictated by the GIL (Global Interpreter Lock) in CPython, which prevents threads from executing Python bytecode in parallel. Process isolation also simplifies state management, as each process has its own address space.

\subsection{Thread Safety by Design}\label{subsec:safety-arch}

In all three applications, thread safety is ensured primarily by architectural means rather than by synchronization primitives: shared data are read-only, and mutable state is isolated per worker.

In \texttt{sim-clone}, the Cherenkov light array (670~MB) is loaded before the parallel section begins and is available to all threads as read-only. Each thread generates photons using its own random number generator and writes the result to a private output buffer. The only aggregation --- computing the global minimum and maximum arrival times --- is performed by the OpenMP runtime via built-in reduction operations, without explicit locks.

In \texttt{sim-trace}, the detector geometry and physics process tables are created by the master thread and are available to workers in read-only mode, as guaranteed by Geant4~MT. All mutable state is encapsulated in a per-worker data structure created at worker initialization. The only mutex --- protecting the file dequeue operation --- is acquired once per event. For a typical event with $\sim 10^5$~photons, the tracing time exceeds the lock acquisition time by orders of magnitude, so contention is negligible.

In \texttt{sim-fit}, process isolation guarantees the absence of shared mutable state by construction. When a shared read-only memory segment is used, the detector pixel coordinates are accessible to all workers but are never modified. The background level is computed once in the master process and passed to workers as an initialization argument.

A summary of the thread-safety mechanisms employed by each application is given in Table~\ref{tab:safety}.

\begin{table}[htbp]
\centering
\caption{Thread-safety mechanisms in the three pipeline applications.}
\label{tab:safety}
\begin{tabular}{lllll}
\hline
Application & Mutexes & Read-only shared data & Per-worker mutable state & Aggregation \\
\hline
sim-clone & 0 & CL array & Per-thread RNG, output buffer & Reduction \\
sim-trace & 1 & Geometry, physics tables & Per-worker event data & Accumulables \\
sim-fit & 0 & Pixel coordinates & Full process state & Async collection \\
\hline
\end{tabular}
\end{table}

\subsection{Physical Correctness under Parallelism}\label{subsec:correctness}

Parallel execution must not violate the statistical properties of the simulation. Correctness is ensured at several levels.

Random number generators are independent across workers. In \texttt{sim-clone}, each thread uses its own generator seeded from a combination of hardware entropy and a hash of the thread identifier, guaranteeing statistical independence of the sequences. In \texttt{sim-trace}, the CLHEP generator is managed by Geant4, which provides per-thread instances. In \texttt{sim-fit}, each process has its own generator state by definition of process isolation.

The simulation result --- the set of approximation parameters --- does not depend on the number of worker threads or processes, up to stochastic fluctuations caused by differences in random number sequences. Deterministic bit-for-bit reproducibility under changes in the degree of parallelism is not guaranteed and is not the goal; only statistical equivalence of the results matters.

The statistical properties of the output sample are likewise preserved. The Poisson distribution of photoelectron counts, the uniform generation of clone positions, and the stochastic photon acceptance all rely on thread-local generators and do not depend on the order in which elements are processed. Permutation of the processing order --- whether of array cells due to dynamic scheduling or of files due to asynchronous collection --- does not affect the statistical properties of the sample.

\subsection{Scalability}\label{subsec:scalability}

The described architecture possesses the following scalability properties:

\begin{itemize}
\item Each stage scales linearly with the number of cores, as a consequence of the absence (or minimality) of inter-thread synchronization. The overhead consists of one-time data loading (the Cherenkov light file, the detector geometry, or the pixel coordinates) and minimal synchronization operations (one mutex acquisition per event in the ray-tracing stage).

\item Scaling with data volume is straightforward: increasing the number of showers, clones, or particle types does not require changes to the architecture. Each new event is processed independently, and adding computational nodes proportionally accelerates the processing.

\item The pipeline is heterogeneous in its use of parallelism technologies (OpenMP, Geant4~MT, multiprocessing), which allows optimal resource utilization at each step. The file system as the interface between stages provides loose coupling: stages can be executed on different computational nodes sharing a common storage.
\end{itemize}

The main scalability limitation is I/O: with a sufficiently large number of threads, performance may be bounded by the file system throughput (especially during the writing of intermediate text files). Switching to binary formats or using a RAM disk can alleviate this limitation if needed.

\section{SPHERE-3 direct CL detector and double registration}\label{sec:double}

We also created a pipeline for simulation of direct CL EAS angular images to be analyzed by the direct CL detector~\cite{some-ref} but its construction is not yet settled so this pipeline is not discussed here. It is subject to a discussion elsewhere.

\begin{figure}[htbp]
\centering
\includegraphics[width=0.8\textwidth]{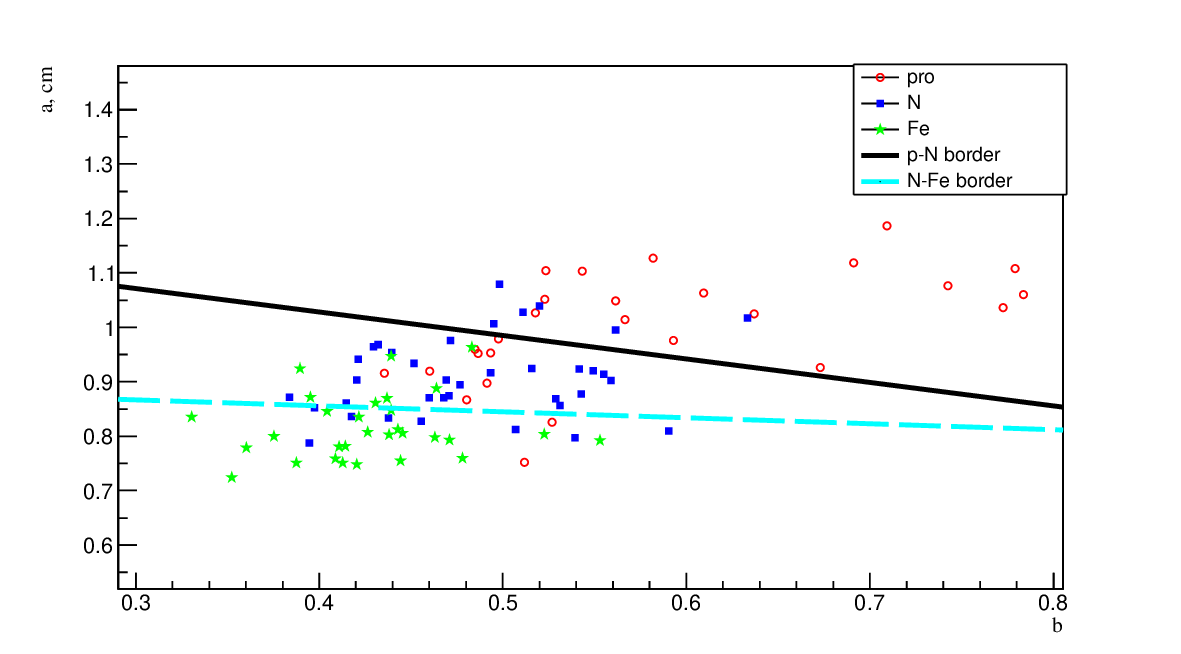}
\caption{Example of joint criteria for event separation by primary mass.
{\it a} is a characteristic of direct CL image, {\it b} is a characteristic of reflected CL image.}
\label{fig:double}
\end{figure}

The power of SPHERE-3 design reveals at full pace when its two detectors work together, i.e. register a shower almost simultaneously and their data are processed as such.
The strategy of the SPHERE-3 experiment intends to maximize the percentage of such double-registration events aiming at substantial increase of the primary parameter estimation accuracy.
Particularly important is the direct CL image sensitivity to the primary mass that exceeds the one of the reflected CL images. We have analysed a number of the direct CL image shape features for such sensitivity~\cite{ICRC} and also developed a few joint (direct+reflected) criteria for EAS separation by mass. Fig.~\ref{fig:double} shows an example of such criteria.

\section{Conclusion}\label{sec:conclusion}

In this work, a software suite with a multi-level parallel architecture for the simulation of the SPHERE-3 experiment is presented --- from the generation of extensive air showers to the approximation of registered Cherenkov light images on the detector mosaic.

The key property of the problem is its natural atomicity: each event at all pipeline stages is processed independently of the others, which allows the computations to scale to an arbitrary number of cores without loss of physical meaning. Each stage employs a parallelism model tailored to the nature of the computational workload: OpenMP for SIMD-like processing of Cherenkov light arrays, Geant4~MT for event-level parallelism during photon tracing, and \texttt{multiprocessing} for process isolation during nonlinear optimization in Python. Thread safety is ensured primarily by architectural means --- shared data are read-only, mutable state is isolated per-worker --- which minimizes synchronization overhead.

The approximation stage implements a robust fitting strategy for the lateral distribution function with multiple restarts and adaptive choice of statistic, which provides stable extraction of shower parameters over a wide range of intensities.

Further development of the suite envisions the integration of the processing stages into a single automated pipeline, as well as the application of the suite to the systematic optimization of the SPHERE-3 detector configuration and the development of criteria for shower classification by primary particle mass.

\begin{acknowledgments}
This work was performed using the equipment of the Shared Research Facilities for Supercomputing at Lomonosov Moscow State University~\cite{SC}. The authors acknowledge support from the Russian Science Foundation (RSF grant No.~23-72-00006). The work of V.A. Ivanov was supported by Theoretical Physics and Mathematics Advancement Foundation ``BASIS'' (grant \#24-2-10-53-1).
\end{acknowledgments}


%
%

\end{document}